\documentstyle[aps,pra]{revtex}
\begin{document}
\draft
\newcommand{\pp}[1]{\phantom{#1}}
\newcommand{\be}{\begin{eqnarray}}
\newcommand{\ee}{\end{eqnarray}}
\newcommand{\ve}{\varepsilon}
\newcommand{\Tr}{{\rm Tr\,}}
\newtheorem{th}{Theorem}
\newtheorem{lem}[th]{Lemma}
%@twocolumnfalse\endcsname

\title{
Simple Proof of Invariance of the Bargmann-Wigner Scalar Products
}
\author{Marek Czachor \cite{*}}
\address{
Wydzia{\l}  Fizyki Technicznej i Matematyki Stosowanej\\
 Politechnika Gda\'{n}ska,
ul. Narutowicza 11/12, 80-952 Gda\'{n}sk, Poland
}
\maketitle
\begin{abstract}
An explicitly covariant formalism for dealing with
Bargmann-Wigner fields is developed. An invariance of the
Barmann-Wigner norm can be proved in a unified way for both
massive and massless fields. It is shown that there exists some
freedom in the choice of the form of the Bargmann-Wigner scalar
product.
\end{abstract}
%\vskip1pc]
%\narrowtext

\section{Introduction}

The main objective of this paper is to
prove invariance of the Bargamann-Wigner scalar
products \cite{BW}
in a manifestly covariant way. There are several
reasons for undertaking this task. One of them is to fill a sort
of gap between the powerful covariant
 spinor methods \cite{PR,PR2} and the noncovariant
methods of  induced representations of the
Poincar\'e group \cite{BR}. The noncovariance
of the induced representations, and
especially of their generators,
manifests itself in  a particular decomposition of
spinor generators into boosts and rotations which is used
to represent boosts by the so-called
Wigner rotations \cite{Ohnuki}. In addition, a
transition between Wigner and spinor bases  involves
dividing a Fourier transform of the spinor field by some powers of
energy $p^0$. Typically, this $p^0$ is identified with the $p^0$
appearing in the invariant measure $d^3p/(2|p^0|)$ and  leads to
the characteristic additional powers $|p^0|^n$ appearing in the
spinor versions of the Bargmann-Wigner products. In the
covariant form of the scalar product given below these
additional powers of energy will be shown to possess some
arbitrariness which is normally hidden behind the noncovariance
of the standard expressions for the scalar products.

Covariant, and especially spinor methods
are known to be a very efficient tool for dealing with
relativistic field theories.
The methods of
Hilbert spaces which, implicitly, are those related to the
Bargmann-Wigner scalar products,
were shown recently to play an important role in
a wavelet formulation of
electrodynamics \cite{K}.
One may hope that the results presented in this paper will prove
useful for the wavelet formulation of higher spin fields.

\section{Massive Bargmann-Wigner Fields}

The Bargmann-Wigner equations \cite{BW,BR,Ohnuki}
representing free spin-$n/2$
fields with mass $m\neq 0$ are equivalent to the set of spinor
field equations for $2^n$ fields $\psi_{A_1\dots A_n}^{0\dots
0}$,  $\psi_{A_1\dots A'_n}^{0\dots 1}$,... $\psi_{A'_1\dots A'_n}^{1\dots
1}$,
\begin{eqnarray}
i\nabla{^A}{_{A'}}\psi(x)^{\dots 0\dots}_{\dots A\dots}
&=&-\frac{m}{\sqrt{2}} \psi(x)^{\dots 1\dots}_{\dots A'\dots},\\
i\nabla{_A}{^{A'}}\psi(x)^{\dots 1\dots}_{\dots A'\dots}
&=&\frac{m}{\sqrt{2}} \psi(x)^{\dots 0\dots}_{\dots A\dots}.
\end{eqnarray}
The convention we use differs slightly from the one introduced
by Penrose and Rindler \cite{PR} (see Appendix~\ref{A2}).
Let $p_\pm^a=(\pm|p^0|,\bbox p)$. The Fourier representation of
the field is
\be
\psi(x)^{\dots}_{\dots}=
\frac{1}{(2\pi)^3}\int\frac{d^3p}{2|p^0|}\,
e^{i\bbox p\cdot\bbox x}
\Bigl\{
e^{-i|p^0|x^0}
\psi_+(\bbox p)^{\dots}_{\dots}
+
e^{i|p^0|x^0}
\psi_-(\bbox p)^{\dots}_{\dots}
\Bigr\}
\ee
where $\psi_\pm(\bbox p)^{\dots}_{\dots}$ satisfy
\begin{eqnarray}
p_\pm{^A}{_{A'}}\psi_\pm(\bbox p)^{\dots 0\dots}_{\dots A\dots}
&=&
-\frac{m}{\sqrt{2}} \psi_\pm(\bbox p)^{\dots 1\dots}_{\dots A'\dots},\\
p_\pm{_A}{^{A'}}\psi_\pm(\bbox p)^{\dots 1\dots}_{\dots A'\dots}
&=&\frac{m}{\sqrt{2}} \psi_\pm(\bbox p)^{\dots 0\dots}_{\dots A\dots}.
\end{eqnarray}
Consider now the tensor
\be
T_\pm(\bbox p)_{a_1\dots a_n}&=&
\psi_\pm(\bbox p)_{A_1\dots A_n}^{0\dots 0}
\bar \psi_\pm(\bbox p)_{A'_1\dots A'_n}^{0\dots 0}
+
\psi_\pm(\bbox p)_{A_1\dots A'_n}^{0\dots 1}
\bar \psi_\pm(\bbox p)_{A'_1\dots A_n}^{0\dots 1}
+\dots
+
\psi_\pm(\bbox p)_{A'_1\dots A'_n}^{1\dots 1}
\bar \psi_\pm(\bbox p)_{A_1\dots A_n}^{1\dots 1}\\
&=&
\psi_\pm(\bbox p)_{A_1\dots A_n}^{0\dots 0}
\overline{\psi_\pm(\bbox p)_{A_1\dots A_n}^{0\dots 0}}
+
\psi_\pm(\bbox p)_{A_1\dots A'_n}^{0\dots 1}
\overline{\psi_\pm(\bbox p)_{A_1\dots A'_n}^{0\dots 1}}
+\dots
+
\psi_\pm(\bbox p)_{A'_1\dots A'_n}^{1\dots 1}
\overline{\psi_\pm(\bbox p)_{A'_1\dots A'_n}^{1\dots 1}}
\ee
where we have, as usual, identified pairs $AA'$ of
spinor indices with the world-vector indices $a$. The standard
Bargmann-Wigner scalar product is defined by the norm
\be
\parallel\psi_\pm\parallel^2=
\int\frac{d^3p}{2|p^0|^{n+1}}\,\bigl\{
\psi_\pm(\bbox p)_{0\dots 0}^{0\dots 0}
\overline{\psi_\pm(\bbox p)_{0\dots 0}^{0\dots 0}}
+
\psi_\pm(\bbox p)_{0\dots 1}^{0\dots 0}
\overline{\psi_\pm(\bbox p)_{0\dots 1}^{0\dots 0}}
+\dots
+
\psi_\pm(\bbox p)_{1'\dots 1'}^{1\dots 1}
\overline{\psi_\pm(\bbox p)_{1'\dots 1'}^{1\dots 1}}\bigr\}
\ee
which being invariant under the Poincar\'e group
is not {\it manifestly\/} invariant. The lack of the manifest
invariance leads to difficulties with applying the
spinor methods in the context of induced representations.

To get the manifestly invariant form we shall first rewrite the
tensor $T_\pm(\bbox p)_{a_1\dots a_n}$ with the help of the
field equations as follows
\be
T_\pm(\bbox p)_{a_1\dots a_k\dots a_n}=
2m^{-2}
p_\pm{_{A_k}}{^{B'_k}}p_\pm{^{B_k}}{_{A'_k}}
T_\pm(\bbox p)_{a_1\dots b_k\dots a_n}=
2m^{-2}
\Bigl(
p_{\pm a_k}p_\pm^{b_k}-\frac{m^2}{2}g{_{a_k}}{^{b_k}}
\Bigr)
T_\pm(\bbox p)_{a_1\dots b_k\dots a_n},
\ee
where we have used the trace-reversal spinor formula \cite{PR}
\be
p{_{A}}{_{B'}}p{_{B}}{_{A'}}=
p_{a}p_{b}-\frac{m^2}{2}g{_{a}}{_{b}}.
\ee
Therefore
\be
T_\pm(\bbox p)_{a_1\dots a_k\dots a_n}=
m^{-2}
p_{\pm a_k}p_\pm^{b_k}
T_\pm(\bbox p)_{a_1\dots b_k\dots a_n}.\label{id}
\ee
Applying (\ref{id}) to itself $n$ times we get
\be
T_\pm(\bbox p)_{a_1\dots a_n}=
m^{-2n}
p_{\pm a_1}\dots p_{\pm a_n}p_\pm^{b_1}\dots p_\pm^{b_n}
T_\pm(\bbox p)_{b_1\dots b_n}.
\ee
The Poincar\'e (i.e. spinor) transformation of the
Bargmann-Wigner field implies
\be
T'_\pm(\bbox p)_{a_1\dots a_n}&=&
m^{-2n}
p_{\pm a_1}\dots p_{\pm a_n}p_\pm^{b_1}\dots p_\pm^{b_n}
T'_\pm(\bbox p)_{b_1\dots b_n}\noindent\\
&=&
m^{-2n}
p_{\pm a_1}\dots p_{\pm a_n}p_\pm^{b_1}\dots p_\pm^{b_n}
\Lambda{_{b_1}}{^{c_1}}\dots\Lambda{_{b_n}}{^{c_n}}
T_\pm(\bbox {\Lambda^{-1}p})_{c_1\dots c_n}\noindent\\
&=&
m^{-2n}
p_{\pm a_1}\dots p_{\pm a_n}
(\Lambda^{-1}p_\pm)^{b_1}
\dots(\Lambda^{-1}p_\pm)^{b_n}
T_\pm(\bbox {\Lambda^{-1}p})_{b_1\dots b_n}
\ee
Let $t_1^{a},\dots,t_n^{a}$ be arbitrary
world-vectors satisfying $t_k^{a}p_{a}\neq 0$ for any $p_a$
belonging to the mass hyperboloid. The expression
\be
\parallel \psi_\pm\parallel'^2=
\int\frac{d^3p}{2|p^0|}
\frac{
t_1^{a_1}\dots t_n^{a_n}
T_\pm(\bbox p)_{a_1\dots a_n}}{
t_1^{b_1}\dots t_n^{b_n}p_{\pm b_1}\dots p_{\pm b_n}}=
m^{-2n}\int\frac{d^3p}{2|p^0|}
p_\pm^{a_1}\dots p_\pm^{a_n}
T_\pm(\bbox p)_{a_1\dots a_n}\label{ex}
\ee
is manifestly invariant. It is interesting that the LHS of
(\ref{ex}) is independent of the choice of
$t_1^{a},\dots,t_n^{a}$ because the RHS does not depend on them.
We can take now $t_k^{a}=t_\pm^{a}$
where $t_\pm^{a}p_{\pm a}$ is equal to $|p^0|$ used in the
invariant measure.  The matrix form of $t_\pm^{AA'}$
is (cf. \cite{PR})
\be
t_\pm^{{\bf AA}'}=t_\pm^{a}g{_a}{^{{\bf AA}'}}=
t_\pm^{0}g{_0}{^{{\bf AA}'}}=
\pm\frac{1}{\sqrt{2}}
\left(
\begin{array}{cc}
1 & 0 \\
0 & 1
\end{array}
\right)
\ee
We have therefore
\be
\parallel \psi_\pm\parallel'^2=(\pm 1)^n%2^{-n/2}
\int\frac{d^3p}{2|p^0|^{n+1}}
T_\pm(\bbox p)_{0\dots 0}=
(\pm 1)^n2^{-n/2}
\parallel \psi_\pm\parallel^2,
\ee
and it follows that the Bargmann-Wigner norm can be written in
the manifestly invariant form
\be
\parallel \psi_\pm\parallel^2=
(\pm 1)^n2^{n/2}m^{-2n}
\int d\mu_m(\bbox p)
p_\pm^{a_1}\dots p_\pm^{a_n}
T_\pm(\bbox p)_{a_1\dots a_n},
\ee
where $d\mu_m(\bbox p)$ is the invariant measure on the mass
hyperboloid. In the simplest example of the Dirac equation
we find
\be
T_\pm(\bbox p)_{a}=g{_a}{^{{AA}'}}\bigl(
\psi_\pm(\bbox p)^0_A\bar \psi_\pm(\bbox p)^0_{A'}+
\psi_\pm(\bbox p)^1_{A'}\bar \psi_\pm(\bbox p)^1_{A}\bigr)=
2^{-1/2}\bar \Psi_\pm(\bbox p)\gamma_a\Psi_\pm(\bbox p),
\ee
where
\be
\Psi_\pm(\bbox p)=
\left(
\begin{array}{c}
\psi_\pm(\bbox p)^0_A\\
\psi_\pm(\bbox p)^1_{A'}
\end{array}
\right)
\ee
is the Dirac bispinor, and
\be
\parallel \psi_\pm\parallel^2=
\pm m^{-2}
\int d\mu_m(\bbox p)
p_\pm^{a}
\bar \Psi_\pm(\bbox p)\gamma_a\Psi_\pm(\bbox p).
\ee
\section{Massless Fields}

A massless spin-$n/2$ field is described by the spinor
equations \cite{PR}
\begin{eqnarray}
\nabla{^{A_k}}{_{A_k'}}\psi(x)_{A_1\dots
A_k\dots A_rA'_1\dots A'_{r\pm n}}&=&0,\label{1}\\
\nabla{_{A_k}}{^{A_k'}}\psi(x)_{A_1\dots A_rA'_1\dots
A'_k\dots A'_{r\pm n}}&=&0\label{2},
\ee
where the spinor $\psi(x)_{A_1\dots A_rA'_1\dots A'_{r\pm n}}$
is totally symmetric in all indices. For simplicity of notation
let us consider the case $r=n$, and a field which has only
unprimed indices.

With any massless field one can associate various types of
potentials \cite{PR2}.
The Hertz-type potentials are defined by
\be
\psi(x)_{A_1\dots  A_n}=\nabla_{A_1A'_1}\dots \nabla_{A_nA'_n}
\xi(x)^{A'_1\dots  A'_n}
\ee
with the subsidiary condition
\be
\Box \xi(x)^{A'_1\dots  A'_n}=0.
\ee
The fact, known generally from the representation theory, that
the field $\psi(x)_{A_1\dots A_n}$ carries only one helicity
(one degree of freedom)
corresponds to the possibility of writing
\be
\xi(x)^{A'_1\dots  A'_n}=\xi^{A'_1\dots  A'_n}\xi(x)
\ee
where $\xi^{A'_1\dots  A'_n}$ is constant and $\Box \xi=0$.

Potentials of another type  are defined by
\be
\psi(x)_{A_1\dots  A_n}=\nabla_{A_1A'_1}\dots \nabla_{A_kA'_k}
\phi(x)^{A'_1\dots  A'_k}_{A_{k+1}\dots A_n},
\ee
and are subject to
\be
\nabla^{A_{k+1}A'_{k+1}}
\phi(x)^{A'_1\dots  A'_k}_{A_{k+1}\dots A_n}=0
\ee
implying the generalized Lorenz gauge
\be
\nabla{^{A_{k+1}}}{_{A'_{k}}}
\phi(x)^{A'_1\dots  A'_k}_{A_{k+1}\dots A_n}=0.
\ee
Let us begin with the Fourier representation of both
 the spinor field
and its Hertz-type potential:
\be
\psi(x)_{A_1\dots A_n}&=&
\frac{1}{(2\pi)^3}\int\frac{d^3p}{2|p^0|}\,
e^{i\bbox p\cdot\bbox x}
\Bigl\{
e^{-i|p^0|x^0}
\psi_+(\bbox p)_{A_1\dots A_n}
+
e^{i|p^0|x^0}
\psi_-(\bbox p)_{A_1\dots A_n}
\Bigr\}\\
\xi(x)^{A'_1\dots A'_n}&=&
\frac{1}{(2\pi)^3}\int\frac{d^3p}{2|p^0|}\,
e^{i\bbox p\cdot\bbox x}
\Bigl\{
e^{-i|p^0|x^0}
\xi_+(\bbox p)^{A'_1\dots A'_n}
+
e^{i|p^0|x^0}
\xi_-(\bbox p)^{A'_1\dots A'_n}
\Bigr\}.
\ee
These definitions imply that
\be
\psi_\pm(\bbox p)_{A_1\dots A_n}=(-i)^n
p_{\pm A_1A'_1}\dots p_{\pm A_nA'_n}
\xi_\pm(\bbox p)^{A'_1\dots A'_n}.\label{pH}
\ee
The rest of the construction is analogous to the massive case.
We define the tensor
\be
T_\pm(\bbox p)_{a_1\dots a_n}&=&
\psi_\pm(\bbox p)_{A_1\dots A_n}
\bar \psi_\pm(\bbox p)_{A'_1\dots A'_n}\\
&=&
p_{\pm A_1B'_1}p_{\pm B_1A'_1}\dots p_{\pm A_nB'_n}p_{\pm B_nA'_n}
\xi_\pm(\bbox p)^{B'_1\dots B'_n}
\bar \xi_\pm(\bbox p)^{B_1\dots B_n}\nonumber\\
&=&
p_{\pm a_1}\dots p_{\pm a_n}p_{\pm b_1}\dots p_{\pm b_n}
U_\pm(\bbox p)^{b_1\dots b_n},
\ee
where
\be
U_\pm(\bbox p)^{b_1\dots b_n}=
\xi_\pm(\bbox p)^{B'_1\dots B'_n}
\bar \xi_\pm(\bbox p)^{B_1\dots B_n}.
\ee
Similarly to the massive case we define
\be
\parallel \psi_\pm\parallel'^2=
\int\frac{d^3p}{2|p^0|}
\frac{
t_1^{a_1}\dots t_n^{a_n}
T_\pm(\bbox p)_{a_1\dots a_n}}{
t_1^{b_1}\dots t_n^{b_n}p_{\pm b_1}\dots p_{\pm b_n}}=
\int\frac{d^3p}{2|p^0|}
p_\pm^{a_1}\dots p_\pm^{a_n}
U_\pm(\bbox p)_{a_1\dots a_n}\label{exx}
\ee
which is manifestly invariant and independent of the choice of
$t_1^{a},\dots,t_n^{a}$.
The expression (\ref{exx}) is directly related to the
Bargmann-Wigner norm. But to see this we first have
to make the one-dimensionality of the representation explicit.

The well known fact that the field
\be
\psi(x)_{A_1\dots A_rA'_1\dots A'_{r\pm n}}
\ee
carries only one helicity can be shown in a covariant manner as
follows. We first contract the field equation (\ref{1})
with $g^a{_{A_kA'_k}}$ and use the identity (cf. Appendix~\ref{A1})
\be
g^a_{\pp A XA'}g^{bYA'}&=&\frac{1}{2}g^{ab}\varepsilon^{{\pp {X}}
Y}_{X}+i{ \sigma}^{ab\pp A
Y}_{\pp {aa}X}
\ee
where ${\sigma}^{ab\pp A Y}_{\pp {aa}X}$ is the generator of
the $(1/2,0)$ spinor representation. Performing an
analogous transformation of
(\ref{2}), denoting $P^a=i\nabla^a$, and introducing the
Pauli-Lubanski tensors corresponding to $(1/2,0)$ and $(0,1/2)$
representations by
\be
S^a{_{X}}{^{Y}}&=&P_b{^*}{\sigma}^{ba\pp A Y}_{\pp {aa}X},\\
S^a{_{X'}}{^{Y'}}&=&P_b{^*}{\bar \sigma}^{ba\pp A Y'}_{\pp
{aa}X'},
\ee
we obtain the equivalent form of (\ref{1}) and (\ref{2})
\be
-\frac{1}{2}P^a\psi(x)_{A_1\dots
A_rA'_1\dots A'_{r\pm n}}&=&S^a{_{A_k}}{^{B_k}}
\psi(x)_{A_1\dots B_k\dots
A_rA'_1\dots A'_{r\pm n}},\label{1'}\\
\frac{1}{2}P^a\psi(x)_{A_1\dots
A_rA'_1\dots A'_{r\pm n}}&=&S^a{_{A'_k}}{^{B'_k}}
\psi(x)_{A_1\dots
A_rA'_1\dots B'_k\dots A'_{r\pm n}}.\label{2'}
\ee
We can further simplify the equations by
introducing the generators
${\sigma}^{ab\pp A {\cal B}}_{\pp {aa}{\cal A}}$
of the $(r/2, r/2\pm n/2)$ representation. With the help of the
respective
Pauli-Lubanski vector the massless equation reduces to
\be
\pm\frac{n}{2}P^a\psi(x)_{\cal A}&=&S^a{_{\cal A}}{^{\cal B}}
\psi(x)_{\cal B},\label{r,s}
\ee
where $\cal A$, $\cal B$  stand for $A_1\dots
A_rA'_1\dots A'_{r\pm n}$, etc.

At the level of the
Fourier transform the one-dimensionality of the representation
follows immediately from the Hertz-type form of the potentials.
Indeed, the momentum representation of the Pauli-Lubanski vector
is
\begin{eqnarray}
-\frac{1}{2}\Bigl(p{_{\pm XA'}}g^{aYA'}-g^a_{\pp A
XA'}p_{\pm}^{YA'}\Bigr)
&=&S^a_\pm(\bbox p){_{X}}{^{Y}},\\
\frac{1}{2}\Bigl(p{_{\pm AX'}}g^{aAY'}-g^a_{\pp A
AX'}p_{\pm}^{AY'}\Bigr)
&=&S^a_\pm(\bbox p){_{X'}}{^{Y'}}.
\end{eqnarray}
Using the trace-reversal formula, the identity
\be
p_{AA'}p^{AB'}=\frac{1}{2}p_a p^a\, \varepsilon{_{A'}}{^{B'}},
\ee
and its complex-conjugated version, we get
\be
S^a_\pm(\bbox p){_{X}}{^{Y}}p{_{\pm YX'}}&=&
-\frac{1}{2}p^a_\pm \,p{_{\pm XX'}},\\
S^a_\pm(\bbox p){_{X'}}{^{Y'}}p{_{\pm XY'}}&=&
\frac{1}{2}p^a_\pm \,p{_{\pm XX'}},
\ee
which imply (\ref{r,s}) which  means
that the spinor
\be
\xi_\pm(\bbox p)^{A'_1\dots A'_n}
\ee
in (\ref{pH}) is in fact arbitrary. The eigenequation
(\ref{r,s}) determines the Fourier components of the field up to
a $\bbox p$-dependent factor (an ``amplitude").
We can write, therefore,
\be
\psi_\pm(\bbox p)_{A_1\dots A_n}=(-i)^n
p_{\pm A_1A'_1}\dots p_{\pm A_nA'_n}
\eta_{\pm}(\bbox p)^{A'_1\dots A'_n}\,f_\pm(\bbox p),\label{pH'}
\ee
where the only restriction on $f_\pm(\bbox p)$ is the
square-integrability of the field, and
$\eta_{\pm}(\bbox p)^{A'_1\dots A'_n}$ is normalized by
\be
p_{\pm b_1}\dots p_{\pm b_n}
\eta_{\pm}(\bbox p)^{B'_1\dots B'_n}
\bar \eta_{\pm}(\bbox p)^{B_1\dots B_n}=(\pm 1)^{n}.
\ee
We can choose $\eta_{\pm}(\bbox p)^{A'_1\dots A'_n}$ as follows.
Let $p_{\pm a}=\pm \pi_{\pm A}\,\bar \pi_{\pm A'}$,
and let $\omega_\pm^{A}$
satisfy $\pi_{\pm A}\omega_\pm^{A}=1$
(i.e. the pair $\pi_{\pm A}$, $\omega_\pm^{A}$
is a spin-frame \cite{PR}). Then
\be
\eta_{\pm}(\bbox p)^{A'_1\dots A'_n}=\bar \omega_\pm^{A'_1}\dots
\bar \omega_\pm^{A'_n},
\ee
and
\be
\psi_\pm(\bbox p)_{A_1\dots A_n}=(\mp i)^n
\pi_{\pm A_1}\dots \pi_{\pm A_n}f_\pm(\bbox p).\label{twist}
\ee
The amplitude then satisfies
\be
(\pm 1)^n\parallel\psi_\pm\parallel'^2=
\int\frac{d^3p}{2|p^0|}\,
|f_\pm(\bbox p)|^2.\label{ibb1}
\ee
Therefore $f_\pm(\bbox p)$ is the Bargmann-Wigner
amplitude which is used in \cite{ibb1,ibb2,ibb3,ibb4} in
the context of the electromagnetic field and the photon wave function.
The form (\ref{twist}) resembles kernels of contour integral
expressions for massless fields arising in the twistor formalism
(cf. \cite{PR2}, Eq.~(6.10.3) on p. 140), and shows that the
Bargmann-Wigner amplitude is closely related to twistor wave
functions.

\section{Acknowledgements}

I would like to thank Prof.~Gerald Kaiser and Prof.~David~E.
Pritchard for their hospitality and support at University of
Massachusetts-Lowell and Massachusetts Institute of Technology
where a part of this work was done. The work was
partly supported by a Fulbright grant.

\section{Appendices}

\subsection{Infeld-van der Waerden tensors and generators of
(1/2,0) and (0,1/2)} \label{A1}

Consider  representations $(\frac{1}{2},0)$ and
 $(0,\frac{1}{2})$
 of an element $\omega\in SL(2,C)$: $e^{\frac{i}{2}
\omega^{ab}{\sigma}_{ab}}$ and
$e^{\frac{i}{2}
 \omega^{ab}\bar {\sigma}_{ab}}$.
 The explicit form of the generators in terms of
Infeld-van der Waerden tensors is
\begin{eqnarray}
\frac{1}{2i}\Bigl(g^a_{\pp A XA'}g^{bYA'}-g^b_{\pp A
XA'}g^{aYA'}\Bigr)
&=&{ \sigma}^{ab\pp A
Y}_{\pp {aa}X},\\
\frac{1}{2i}\Bigl(
g^a_{\pp A AX'}g^{bAY'}-g^b_{\pp A AX'}g^{aAY'}\Bigr)&=&\bar
{ \sigma}^{ab\pp A
Y'}_{\pp {aa}X'}.
\end{eqnarray}
Their purely spinor form is
\begin{eqnarray}
{\sigma}_{AA'BB'XY}&=&\frac{1}{2i}\varepsilon
_{A'B'}\bigl(\varepsilon_{AX}\varepsilon_{BY}+
\varepsilon_{BX}\varepsilon_{AY}\bigr),\\
\bar {\sigma}_{AA'BB'X'Y'}
&=&\frac{1}{2i}\varepsilon_{AB}\bigl(\varepsilon_{A'X'}\varepsilon_{B'Y'}+
\varepsilon_{B'X'}\varepsilon_{A'Y'}\bigr),
\end{eqnarray}
Dual tensors are $^*\bar
{\sigma}^{ab\pp A
Y'}_{\pp {aa}X'}=+i\bar
{\sigma}^{ab\pp A
Y'}_{\pp {aa}X'}$
and
$^*{\sigma}^{ab\pp A
Y}_{\pp {aa}X}=-i{\sigma}^{ab\pp A
Y}_{\pp {aa}X}$.

Additionally the Infeld-van der Waerden tensors satisfy
\begin{eqnarray}
g^a_{\pp A XA'}g^{bYA'}+g^b_{\pp A
XA'}g^{aYA'}
&=&g^{ab}\varepsilon^{{\pp {X}}
Y}_{X}\label{IW1}\\
g^a_{\pp A AX'}g^{bAY'}+g^b_{\pp A
AX'}g^{aAY'}
&=&g^{ab}\varepsilon^{{\pp {X}}
Y'}_{X'}\label{IW2}
\end{eqnarray}
These equations lead to the useful expressions
\be
g^a_{\pp A XA'}g^{bYA'}&=&\frac{1}{2}g^{ab}\varepsilon^{{\pp {X}}
Y}_{X}+i{ \sigma}^{ab\pp A
Y}_{\pp {aa}X}\\
g^a_{\pp A AX'}g^{bAY'}&=&\frac{1}{2}g^{ab}\varepsilon^{{\pp {X}}
Y'}_{X'}+i
\bar
{ \sigma}^{ab\pp A
Y'}_{\pp {aa}X'}
\ee

\subsection{Spinor and bispinor forms of the Dirac equation}
\label{A2}

The matrix form of the Dirac in the momentum representation
equation can be written explicitly  as
\be
\left(
\begin{array}{cc}
0 & (p^0 + \bbox p\cdot\bbox \sigma){^A}{^{B'}}\\
(p^0 - \bbox p\cdot\bbox \sigma){_{A'}}{_{B}} & 0
\end{array}
\right)
\left(
\begin{array}{c}
\psi^B\\
\xi_{B'}
\end{array}
\right)
=
\left(
\begin{array}{cc}
0 & p^a\sigma_a{^A}{^{B'}}\\
p^a\tilde \sigma_a{_{A'}}{_{B}} & 0
\end{array}
\right)
\left(
\begin{array}{c}
\psi^B\\
\xi_{B'}
\end{array}
\right)
=m
\left(
\begin{array}{c}
\psi^A\\
\xi_{A'}
\end{array}
\right),
\ee
where
\be
\sigma_a{^A}{^{B'}}&=&(\bbox 1, \bbox \sigma){^A}{^{B'}}\\
\tilde \sigma_a{_{A'}}{_{B}}&=&(\bbox 1,- \bbox \sigma){_{A'}}{_{B}},
\ee
and $\bbox \sigma$ is a matrix vector whose components are the
Pauli matrices.
The matrix formulas
\be
\tilde \sigma_a \sigma_b+ \tilde \sigma_b \sigma_a &=&
2g_{ab}\bbox 1\\
 \sigma_a \tilde \sigma_b+ \sigma_b\tilde \sigma_a &=&
2g_{ab}\bbox 1
\ee
have the following spinor form
\be
\tilde \sigma_a{_{A'}}{_{X}} \sigma_b{^X}{^{B'}}+ \tilde
\sigma_b{_{A'}}{_{X}} \sigma_a{^X}{^{B'}} &=&
2g_{ab}\varepsilon{_{A'}}{^{B'}}\\
 \sigma_a{^B}{^{X'}} \tilde \sigma_b{_{X'}}{_{A}}+
\sigma_b{^B}{^{X'}}\tilde \sigma_a
{_{X'}}{_{A}} &=&
2g_{ab}\varepsilon{_A}{^{B}}
\ee
which compared with (\ref{IW1}), (\ref{IW2})
shows that
\be
g_a{_{A}}{_{B'}}&=&\frac{1}{\sqrt{2}}\tilde
\sigma_a{_{B'}}{_{A}}\\
g_a{^A}{^{B'}}&=&\frac{1}{\sqrt{2}}
\sigma_a{^A}{^{B'}}
\ee
The Dirac equation in the Minkowski representation is ($\hbar=1$)
\begin{eqnarray}
i\nabla_{AA'}\psi^A&=&\frac{m}{\sqrt{2}} \xi_{A'},\\
i\nabla^{AA'}\xi_{A'}&=&\frac{m}{\sqrt{2}} \psi^{A}
\end{eqnarray}
where $\nabla_{AA'}=\nabla^ag_{aAA'}$ etc. This equation differs
by a sign and the presence of $i$ from the
form given in \cite{PR}.
The matrix form of the equation
\be
p^q\left(
\begin{array}{cc}
0 & g{_{qA}}{^{B'}}\\
-g{_q}{^{B}}{_{A'}} & 0
\end{array}
\right)
\left(
\begin{array}{c}
\psi_B\\
\xi_{B'}
\end{array}
\right)
=\frac{m}{\sqrt{2}}
\left(
\begin{array}{c}
\psi_A\\
\xi_{A'}
\end{array}
\right)
\ee
shows that the Dirac gamma matrices are given by
\be
\gamma{_q}{_\alpha}{^\beta}=
\sqrt{2}\left(
\begin{array}{cc}
0 & g{_{qA}}{^{B'}}\\
-g{_q}{^B}{_{A'}} & 0
\end{array}
\right)
\ee
Product of two gamma matrices
\be
\gamma{_q}{_\alpha}{^\beta}\gamma{_r}{_\beta}{^\gamma}&=&
\left(
\begin{array}{cc}
g{_{qr}}\varepsilon{_A}{^C}+2i\sigma{_{qr}} {_A}{^C} & 0\\
0 & g{_{qr}}\varepsilon{_{A'}}{^{C'}}+2i\bar \sigma{_{qr}}
{_{A'}}{^{C'}}
\end{array}
\right)=
g{_{qr}}I{_\alpha}{^\gamma}+2i\sigma{_{qr}} {_\alpha}{^\gamma}
\ee
implies
\be
\gamma{_q}{_\alpha}{^\beta}\gamma{_r}{_\beta}{^\gamma}+
\gamma{_r}{_\alpha}{^\beta}\gamma{_q}{_\beta}{^\gamma}
&=&2g{_{qr}}I{_\alpha}{^\gamma}\label{znak1},\\
\gamma{_q}{_\alpha}{^\beta}\gamma{_r}{_\beta}{^\gamma}-
\gamma{_r}{_\alpha}{^\beta}\gamma{_q}{_\beta}{^\gamma}
&=&4i\sigma{_{qr}} {_\alpha}{^\gamma}.\label{znak2}
\ee
(\ref{znak2}) differs by the factor $(-1/2)$
from the definition from \cite{BD}
 because there the  generators are defined
by $S(\omega)=e^{-\frac{i}{4}
\omega^{ab}{\sigma}_{ab}}$. There is also a difference with
respect to \cite{PR} where the gamma matrices
are defined
 without the $-$
sign  (this would
lead to the opposite sign at the RHS of (\ref{znak1})).

The spinor form of the Dirac current is
\be
j_a=\sqrt{2}g_a{^{AA'}}\bigl(\psi_A\bar \psi_{A'} +
\xi_{A'}\bar \xi_{A}\bigr).\label{spin-cur}
\ee
(\ref{spin-cur}) is derived spinorially as follows
\be
j_a=
\sqrt{2}(\bar \psi^{A'},\bar \xi^A)
\left(
\begin{array}{cc}
0 & \varepsilon{_{A'}}{^{B'}}\\
-\varepsilon{_A}{^{B}} & 0
\end{array}
\right)
\left(
\begin{array}{cc}
0 & g{_{aB}}{^{C'}}\\
-g{_a}{^C}{_{B'}} & 0
\end{array}
\right)
\left(
\begin{array}{c}
\psi_C\\
\xi_{C'}
\end{array}
\right)
\ee
and
\be
j_0&=&
(\bar \psi^{A'},\bar \xi^A)
\underbrace{
\left(
\begin{array}{cc}
0 & \varepsilon{_{A'}}{^{B'}}\\
-\varepsilon{_A}{^{B}} & 0
\end{array}
\right)}_{``\gamma_0"}
\underbrace{\sqrt{2}
\left(
\begin{array}{cc}
0 & g{_{0B}}{^{C'}}\\
-g{_0}{^C}{_{B'}} & 0
\end{array}
\right)}_{``\gamma_0"}
\left(
\begin{array}{c}
\psi_C\\
\xi_{C'}
\end{array}
\right)\nonumber\\
&=&
\sqrt{2}
(\bar \psi_{A'},\bar \xi_A)
\left(
\begin{array}{cc}
g{_0}{^A}{^{A'}} & 0\\
0 & g{_0}{^A}{^{A'}}
\end{array}
\right)
\left(
\begin{array}{c}
\psi_A\\
\xi_{A'}
\end{array}
\right)
\ee
showing that the matrix $\gamma_0$ appearing in textbooks
corresponds actually to two different spinor objects.
The pseudoscalar matrix $\gamma_5$ corresponds to the spinor
matrix
\be
\gamma{_5}{_X}{^Y}&=&
\frac{i}{4!}
e^{abcd}\gamma_a\gamma_b\gamma_c\gamma_d{_X}{^Y}
=
\left(
\begin{array}{cc}
-\ve{_X}{^Y} & 0\\
0 &  \ve{_{X'}}{^{Y'}}
\end{array}
\right).
\ee

\subsection{Alternative covariant proof for the Maxwell field}
\label{A3}

The other form of potentials
is not very helpful in proving invariance of the Bargmann-Wigner
norm in the general spin case. It is instructive, however, to see
how the spinor language simplifies the standard
proof in the particular case of the Maxwell field (cf. \cite{G}
and \cite{K}).

Consider the electromagnetic spinor
\be
\varphi_{\pm}(\bbox p)_{AB}=\frac{i}{2}F_\pm^{qr}(\bbox p)
\sigma_{qrAB},
\ee
which satisfies
\be
\varphi_{\pm}(\bbox p)_{AB}=-ip_{\pm AA'}\phi_\pm(\bbox p){_{B}}{^{A'}}
=-ip_{\pm BA'}\phi_\pm(\bbox p){_{A}}{^{A'}}
\ee
implying the Lorenz gauge
\be
p_{\pm AA'}\phi_\pm(\bbox p){^{AA'}}=0
\ee
for the 4-vector potential $\phi_\pm^a(\bbox p)$.
We consider the tensor
\be
T_{\pm ab}(\bbox p)&=&
\varphi_\pm(\bbox p)_{AB}\bar \varphi_\pm(\bbox p)_{A'B'}=
p_{\pm AC'}\phi_\pm(\bbox p){_{B}}{^{C'}}
p_{\pm CA'}\phi_\pm(\bbox p){^{C}}{_{B'}}
=
p_{\pm AA'}p_{\pm CC'}\phi_\pm(\bbox p){_{B}}{^{C'}}
\phi_\pm(\bbox p){^{C}}{_{B'}}\nonumber\\
&=&
p_{\pm AA'}p_{\pm BC'}\phi_\pm(\bbox p){_{C}}{^{C'}}
\phi_\pm(\bbox p){^{C}}{_{B'}}=
-\frac{1}{2}p_{\pm a}p_{\pm b}\phi{_{\pm c}}(\bbox p)
\phi{_{\pm}^c}(\bbox p),
\ee
where we have used the trace-reversal identity and the fact that
\be
\phi_\pm(\bbox p){_{CC'}}
\phi_\pm(\bbox p){^{C}}{_{B'}}=
-\phi_\pm(\bbox p){_{CB'}}
\phi_\pm(\bbox p){^{C}}{_{C'}}.
\ee
The tensor satisfies the formula
\be
T_{\pm ab}(\bbox p)&=&
\frac{1}{2}\Bigl(
\frac{1}{4}g_{ab}F_{\pm cd}(\bbox p)F_\pm^{cd}(\bbox p)
-F_{\pm ac}(\bbox p)F{_{\pm b}}^{c}(\bbox p)\Bigr),
\ee
and, in particular,
\be
T_{\pm 00}(\bbox p)&=&
\frac{1}{4}\Bigl(
\bbox E_\pm(\bbox p)^2 + \bbox B_\pm(\bbox p)^2\Bigr),
\ee
where $\bbox E_\pm(\bbox p)$ and $\bbox B_\pm(\bbox p)$ are the
positive and negative frequency Fourier transforms of the
electromagnetic field.

Now we can repeat the reasoning presented above for the general
case and the norm used in the wavelet analysis of the
electromagnetic field \cite{K} becomes a particular case of
\be
\parallel \varphi\parallel^2=
\parallel \varphi_+\parallel'^2
+
\parallel\varphi_-\parallel'^2,
\ee
where
\be
\parallel \varphi_\pm\parallel'^2=
\int\frac{d^3p}{2|p^0|}
\frac{
t_1^{a}t_2^{b}
T_{\pm ab}(\bbox p)
}{
t_1^{c} t_2^{d}p_{c} p_{d}}.
\ee


\begin{references}
\bibitem[*]{*}Electronic address: mczachor@sunrise.pg.gda.pl

\bibitem{BW}V.~Bargmann and E.~P.~Wigner, Proc. Nat. Acad. Sci.
USA {\bf 34}, 211 (1948).
\bibitem{PR}R.~Penrose and W.~Rindler, {\it Spinors and
Space-Time\/}, vol.~1 (Cambridge University Press, 1984).
\bibitem{PR2}R.~Penrose and W.~Rindler, {\it Spinors and
Space-Time\/}, vol.~2 (Cambridge University Press, 1986).
\bibitem{BR}A.~O.~Barut and R.~Raczka, {\it Theory of Group
Representations and Applications\/} (Polish Scientific
Publishers, Warszawa, 1980).
\bibitem{Ohnuki}Y.~Ohnuki, {\it Unitary Representations of the
Poincar\'e Group and Relativistic Wave Equations\/} (World
Scientific, Singapore, 1988).
\bibitem{K}G.~Kaiser, {\it A Friendly Guide to Wavelets\/}
(Birkh\"auser, Boston, 1994).
\bibitem{ibb1}I.~Bia{\l}ynicki-Birula and
Z.~Bia{\l}ynicka-Birula, {\it Quantum Electrodynamics\/}
(Pergamon, Oxford, 1976).
\bibitem{ibb2}I.~Bia{\l}ynicki-Birula and
Z.~Bia{\l}ynicka-Birula, Phys.~Rev.~D {\bf 35}, 2383 (1987).
\bibitem{ibb3}I.~Bia{\l}ynicki-Birula, Acta Phys. Polon. A {\bf
86}, 97 (1994).
\bibitem{ibb4}I.~Bia{\l}ynicki-Birula, in {\it Coherence and
Quantum Optics VII --- Rochester '95\/}, Plenum Press, to be
published.
\bibitem{BD}J.~D.~Bjorken and S.~D.~Drell, {\it Relativistic
Quantum Mechanics\/} (McGraw-Hill, 1964).
\bibitem{G}L.~Gross, J.~Math.~Phys. {\bf 5}, 687 (1964).
\end{references}
\end{document}